\documentclass[10pt,journal]{IEEEtran}

\usepackage{balance,subscript,gensymb,amsmath,graphicx}

\newcommand{\lo}[1]{\textsubscript{#1}}
\newcommand{\hi}[1]{\textsuperscript{#1}}

\ifCLASSINFOpdf
\else
\fi
\begin{document}
%
\title{Effects of Low Temperature Annealing on the Transport Properties of Zinc Tin Nitride}
%
%
%

\author{Angela~N.~Fioretti,~\IEEEmembership{Member,~IEEE,}
        Eric~S.~Toberer,~\IEEEmembership{Member,~IEEE}
        Andriy~Zakutayev,~\IEEEmembership{Member,~IEEE,}
        and~Adele~C.~Tamboli,~\IEEEmembership{Member,~IEEE}
\thanks{A. N. Fioretti is with the National Renewable Energy Laboratory and the Colorado School of Mines, Golden, CO, 80401 USA e-mail: afiorett@mines.edu}
\thanks{E. S. Toberer and A. C. Tamboli are with the National Renewable Energy Laboratory and the Colorado School of Mines, Golden, CO, 80401 USA}
\thanks{A. Zakutayev is with the National Renewable Energy Laboratory, Golden, CO 80401 USA}%
\thanks{Manuscript received June 18, 2015; revised Month Day, 2015.}}

%
%

\markboth{Journal of Photovoltaics,~Vol.~XX, No.~X, June~2015}%
{Fioretti \MakeLowercase{\textit{et al.}}: Effects of Low Temperature Annealing on the Transport Properties of Zinc Tin Nitride}
%



\maketitle

\begin{abstract}
ZnSnN\lo{2} has recently garnered increasing interest as a potential solar absorber material due to its direct bandgap that is predicted to be tunable from 1.0-2.1 eV based on cation disorder. One important challenge to the further development of this material for photovoltaics (PV) is to reliably synthesize films with carrier density $\leq$10\hi{17} electrons cm\hi{-3}. In this work, we perform a systematic annealing study on compositionally-graded Zn-Sn-N thin films to determine the effects on carrier density and transport of such post-growth treatment. We find that annealing up to 6 hr under an activated nitrogen atmosphere results in a reduction in carrier density by $\sim$80\% for zinc-rich films, and by $\sim$50\% for stoichiometric films. However, we also find that annealing reduces mobility as a function of increasing annealing time. This result suggests that initial film disorder hampers the benefits to film quality that should have been gained through annealing. This finding highlights that carefully managed initial growth conditions will be necessary to obtain PV-quality ZnSnN\lo{2} absorber films.
\end{abstract}


%
\IEEEpeerreviewmaketitle

\section{Introduction}
\label{sec:intro}
%
%
%
%
\IEEEPARstart{T}{oday's} landscape of photovoltaic (PV) absorbers is experiencing a renaissance in new materials brought on both by advances in predictive computational methods and in high throughput materials assessment \cite{zhang2012,zakutayev2013c}. Entirely new classes of absorbers materials are currently being realized, each with their own potential benefits for solar energy technology. Within this host of novel materials, one evident choice for long-term terawatt scale generation is that of novel Earth-abundant absorbers. Many of the materials within this category have been predicted to possess highly desirable qualities for PV, and yet remain unexplored or under-researched \cite{yu2013,zakutayev2014,zawadzki2013}. One such material is ZnSnN\lo{2}; an Earth-abundant II-IV-V\lo{2} semiconductor with a low electron effective mass and a direct bandgap between 1.0--2.1 eV \cite{quayle2013,lahourcade2013,feldberg2013}.

Within recent years, ZnSnN\lo{2} has become the topic of ever-increasing research interest. This interest is due in large part to its direct gap that is possibly tunable based on degree of cation disorder \cite{chen2014}. This range of possible bandgap energies makes ZnSnN\lo{2} a strong potential candidate for single or multi-junction solar cells depending upon the bandgap value accessed. Reports of photoluminescence response at both low temperature \cite{fioretti2015,quayle2015} and room temperature \cite{quayle2013}, and reduced carrier density  \cite{fioretti2015,deng2015}, categorize ZnSnN\lo{2} as a possible absorber material for PV. In addition to this list of desirable traits, ZnSnN\lo{2} is also a wurtzite-based, tetrahedrally-bonded nitride, and is therefore expected to exhibit good stability.

One main challenge affecting ZnSnN\lo{2} development is the ability to lower carrier density beyond the 10\hi{17}--10\hi{18} range that is currently achievable \cite{fioretti2015,deng2015}, while also maintaining mobility at the currently reported 10 cm\hi{2} V\hi{-1} s\hi{-1} \cite{feldberg2013,lahourcade2013}. An equally important challenge is for the field of ZnSnN\lo{2} research to converge on a bandgap value for cation-ordered structure. In the present work, we sought to address the former challenge by performing a systematic annealing study on compositionally-graded Zn-Sn-N thin films to determine the effect of annealing time at fixed temperature on carrier density and mobility. We find, quite non-intuitively, that longer annealing time results in consistently decreasing mobility, suggesting that initial film disorder hampers the benefits that might have been gained through post-growth anneals. We also find that zinc-rich films exhibit the largest percent decrease in free electron concentration of any of the film compositions explored, and that anneal times less than 6 hr are optimal for such films based on the observation of increasing carrier density with longer annealing treatment. Finally, we determine that annealing temperature above 300\degree C may be necessary to induce equiaxed grain structure and observe the effects on absorption onset of super-lattice ordering. Collectively, these findings highlight the importance of carefully managed growth conditions, rather than relying on post-growth annealing, in order to obtain PV-quality ZnSnN\lo{2} absorber films.

\section{Experimental Methods}
\label{sec:exp}

	\subsection{Thin Film Synthesis and Annealing}
Compositionally-graded thin films libraries were deposited isothermally on glass substrates heated to 230\degree C using combinatorial RF-cosputtering as described previously \cite{fioretti2015}. Further information regarding our implementation of the combinatorial approach can be found elsewhere \cite{caskey2014,subramaniyan2014,zakutayev2013}. Four such libraries were prepared and subjected to annealing at 300\degree C for 0 hr, 3 hr, 6 hr, and 14 hr, respectively. Annealing was performed under an activated nitrogen atmosphere provided by an RF nitrogen atomic source and at a pressure of 20 mTorr in the same vacuum chamber in which films were grown, without breaking vacuum between growth and anneal. The sample that was not annealed was used as a reference for comparison to the annealed samples. Samples described in this work were typically 400--500~nm thick.

	\subsection{Characterization}
Mapping-style characterization was used to collect four-point probe (4pp) sheet resistance, X-ray diffraction (XRD), X-ray fluorescence (XRF), and UV-Vis-NIR spectroscopy data. More details on the combinatorial characterization method used in this work have been published previously \cite{fioretti2015}. XRF was used to obtain composition data in the form of zinc and tin atomic percents, which were subsequently used to calculate the Zn/(Zn+Sn) ratios referred to throughout this work. To obtain the thickness data used to calculate absorption coefficient and conductivity, 3--5 representative pieces $\sim$7~x~7~mm in size were cut from each library spanning the composition gradient for measurement with scanning electron microscopy (SEM). Hall effect measurements were used to collect carrier density and mobility data, using the thickness values determined by SEM. Eight samples   of the 7~x~7~mm size and spanning the full composition gradient for each library were used for performing Hall effect. Contacts for Hall effect were made with soldered indium. More details on the non-mapping characterization methods described here are available from Ref. \cite{fioretti2015}.

\begin{figure}[!t]
	\centering
		\includegraphics[width=8cm]{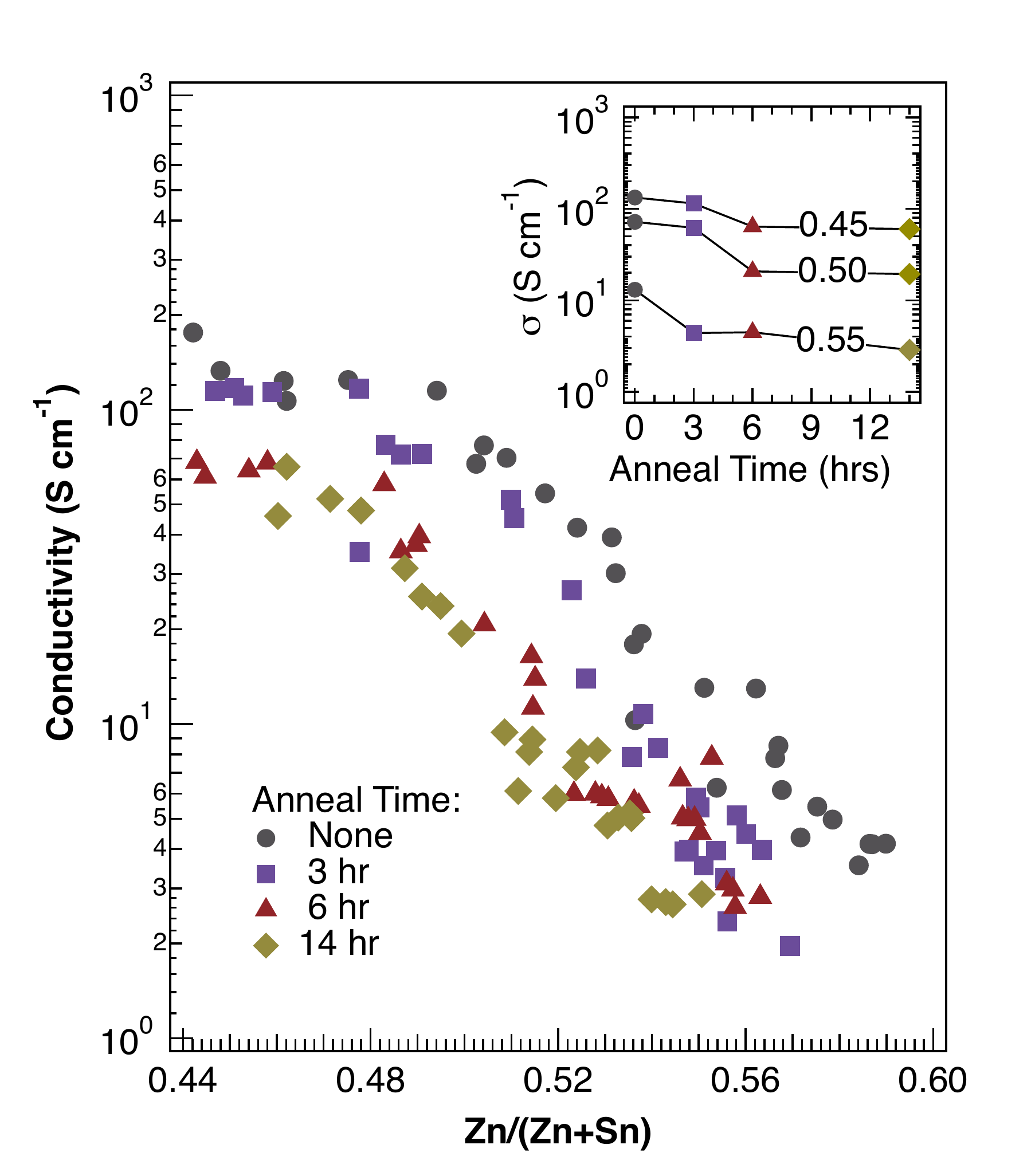}
	\caption{Conductivity as measured by four-point probe plotted as a function of zinc fraction on the cation site, with color indicating annealing time. The inset shows the decreasing trend in conductivity with increased annealing time for three particular cation compositions. In general, annealing treatment longer than 6 hr did not yield additional decreases in conductivity.}
	\label{fig:sigma}
\end{figure} 

\section{Results and Discussion}
\label{sec:results}

\subsection{Electrical and Transport Properties}
\label{sec:elect}

The conductivity of films annealed under activated nitrogen was found to decrease relative to as-deposited films as a function of annealing time at 300\degree C (Fig.~\ref{fig:sigma}). This decrease in conductivity was most pronounced for tin-rich and stoichiometric films after a 6 hr annealing treatment, and for zinc-rich films after a 3 hr treatment. No further decrease in conductivity was observed for anneal time greater than 6 hr for either tin-rich or stoichiometric films. For zinc-rich films, no appreciable decrease in conductivity was observed beyond 3 hr of anneal time. Decreasing conductivity with anneal time may appear desirable, given that such a trend could indicate decreased carrier density. However, the transport properties determined by Hall effect (Figs.~2a-b, discussed below) reveal a consistent decrease in mobility with annealing time in addition to changes in carrier density; a trend which dominated the decreasing conductivity observed in Fig.~\ref{fig:sigma}. 

We note, parenthetically, that the more obvious downward trend in conductivity observed in Fig. \ref{fig:sigma} occurs as a function of zinc fraction on the cation site (i.e. Zn/(Zn+Sn)), with zinc-rich compositions exhibiting lower conductivity compared to tin-rich compositions. This phenomenon has been reported in our previous work on doping control and transport in the Zn-Sn-N material system \cite{fioretti2015}. As described therein, this relationship between lower conductivity and higher zinc content is most likely due to the formation of defect complexes that compensate for anion-site defects, such as O\lo{N} substitutions \cite{fioretti2015}.

\begin{figure}[!t]
	\centering
		\includegraphics[width=5cm]{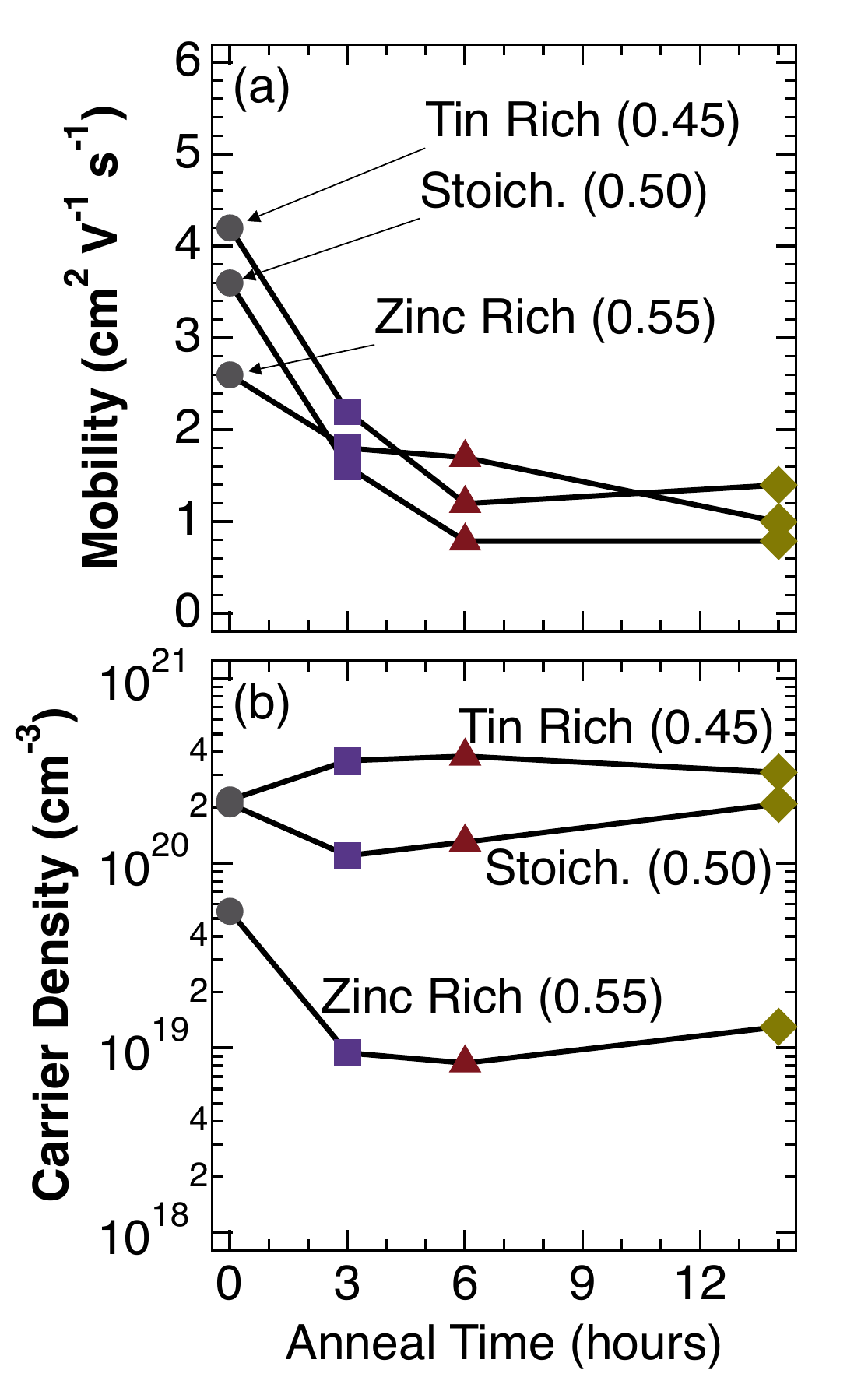}
	\caption{Mobility (a) and carrier density (b) as a function of annealing time for three values of Zn/(Zn+Sn). Mobility was found to consistently decrease upon annealing treatment, likely due to clustering of non-octet-rule-preserving motifs. Zinc-rich films exhibited the largest percent decrease in carrier density overall, which was observed after a 3 hr annealing treatment.}
	\label{fig:trans}
\end{figure} 

As mentioned above, post-growth annealing treatments consistently lowered the mobility for all films described in this work (Fig. \ref{fig:trans}a). Although this result is notably non-intuitive from a basic materials science standpoint, there is some precedence in the literature for observing such an effect. A previous work on annealing ZnSnN\lo{2} thin films reported decreased mobility values as annealing temperature increased from 300\degree C to 400\degree C \cite{deng2015}. One possible explanation for lowered mobility upon annealing in ZnSnN\lo{2} can be found in a recent computational work on compositional inhomogeneity in multinary, tetrahedrally-bonded semiconductors \cite{zawadzki2015}. From that work, compositional inhomogeneity essentially refers to the entropy-driven tendency for non-octet-rule-preserving motifs (tetrahedra) to cluster together upon heat treatment or during high temperature growth. Such clustering was shown to result in charge localization, thus increasing the concentration of charged scattering centers and subsequently decreasing the mobility. By annealing sputtered films of ZnSnN\lo{2} that undoubtedly possessed a considerable degree of cation disorder to begin with, this work and the previous one may have simply encouraged motif clustering, thus lowering mobility. From this perspective, one potential strategy to avoid motif clustering while still being able to improve material properties through annealing would be to begin with films grown under conditions that encourage increased order, such as growing films on lattice-matched substrates.

\begin{figure}[!t]
	\centering
		\includegraphics[width=7cm]{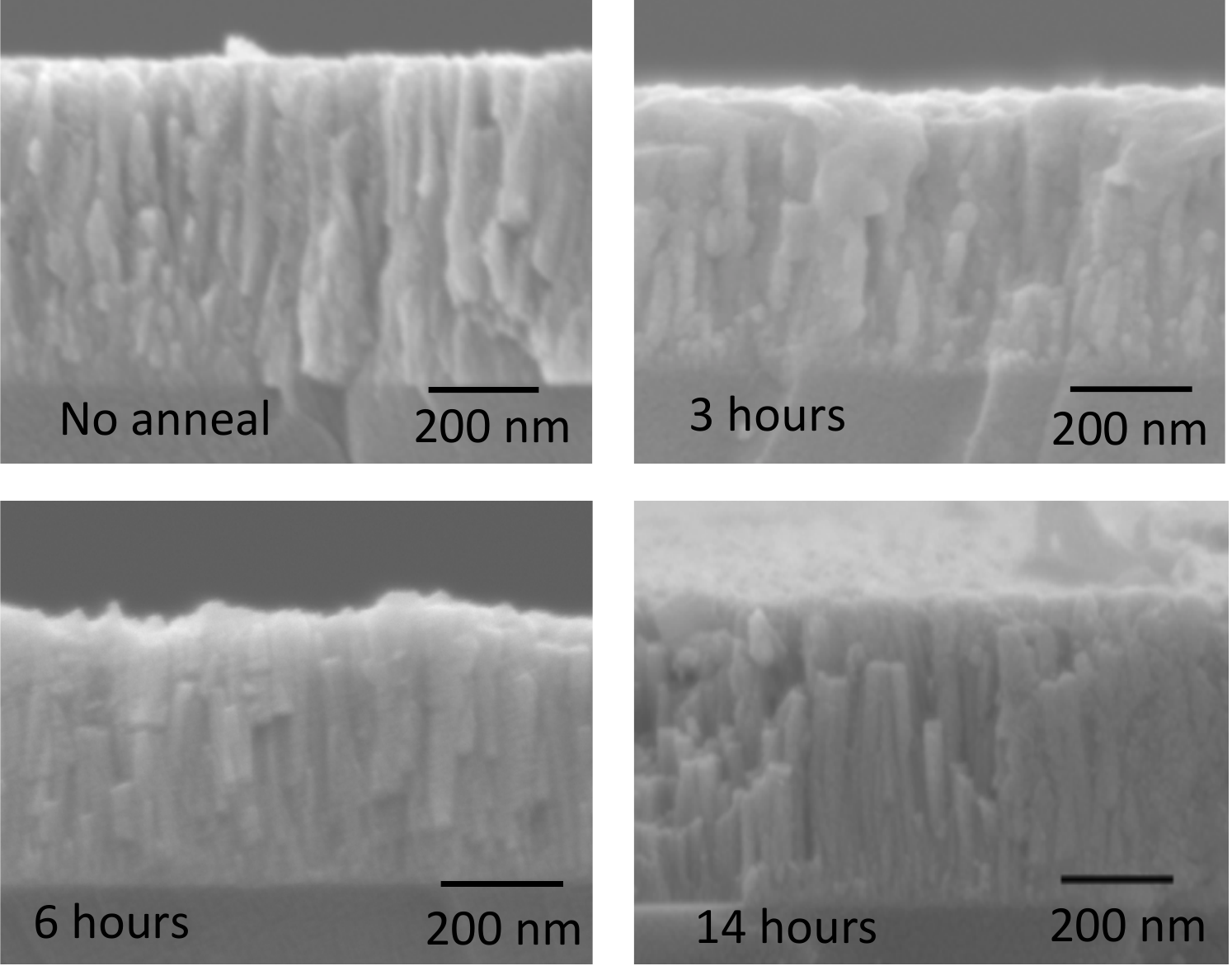}
	\caption{Cross-sectional SEM images of different ZnSnN\lo{2} films annealed for various times at 300\degree C. Columnar grain structure was observed for all films described in this work no matter the annealing time, and little if any grain coalescence could be observed by SEM.}
	\label{fig:sem}
\end{figure} 

Carrier density for films in this work exhibited the largest relative decrease for stoichiometric and zinc-rich films after 3 hr of annealing time, and then increased again after a 14 hr anneal. The opposite effect on carrier density was observed for tin-rich films as annealing time increased. Carrier density as a function of annealing time for tin-rich, stoichiometric, and zinc-rich films is shown in Fig. \ref{fig:trans}b. Given that tin-rich Zn-Sn-N films in this work and our previous work reliably exhibited carrier density \textgreater10\hi{20} electrons/cm\hi{3}, the finding that annealing such films increases the electron concentration effectively rules out tin-rich Zn-Sn-N as a candidate absorber material for PV \cite{fioretti2015}. On the other hand, zinc-rich films consistently exhibited the lowest carrier density overall in this work and our previous on Zn-Sn-N, and additionally showed the largest percent decrease in carrier density with annealing in Fig. \ref{fig:trans}b. The dip in electron concentration for stoichiometric and zinc-rich films after a 3 hr anneal followed by rising electron concentration after longer anneal times suggests two different effects may be occurring. It is possible that short anneal times facilitate defect complex formation (thus lowering carrier density), but that annealing treatments longer than 3 hr leads to the formation of some other donor defect. This observation suggests that shorter annealing treatments of 6 hr or less are better suited to improving film quality for PV absorber applications.

\subsection{Structural Properties and Light Absorption}
\label{sec:struclight}

We now turn our attention to the effects of annealing on the morphology and structure of stoichiometric Zn-Sn-N films. Fig. \ref{fig:sem} shows cross-sectional images of nominally stoichiometric films grown at 230\degree C and annealed at 300\degree C for various periods of time. Columnar growth was observed for all films described in this work no matter the anneal time, with minimal if any grain coalescence taking place as annealing time increased. Grain size for the films described in this work was found to be $\sim$30 nm on average, as determined by SEM. This finding suggests that an annealing temperature of 300\degree C did not provide enough thermal energy for grain growth (a.k.a Ostwald ripening) in the amount of time given. Considering that the carrier density (Fig. \ref{fig:trans}) reached a minimum after only 3--6 hr of annealing time and that mobility continued to decrease with longer annealing, it would not be sensible (nor practical) to anneal longer than 14 hr to obtain the desired grain coalescence. Therefore, we conclude that higher annealing temperature would be the best route forward to obtain equi-axed grains while also lowering carrier density, if non-epitaxial growth is to be pursued. 

\begin{figure}[!t]
	\centering
		\includegraphics[width=6cm]{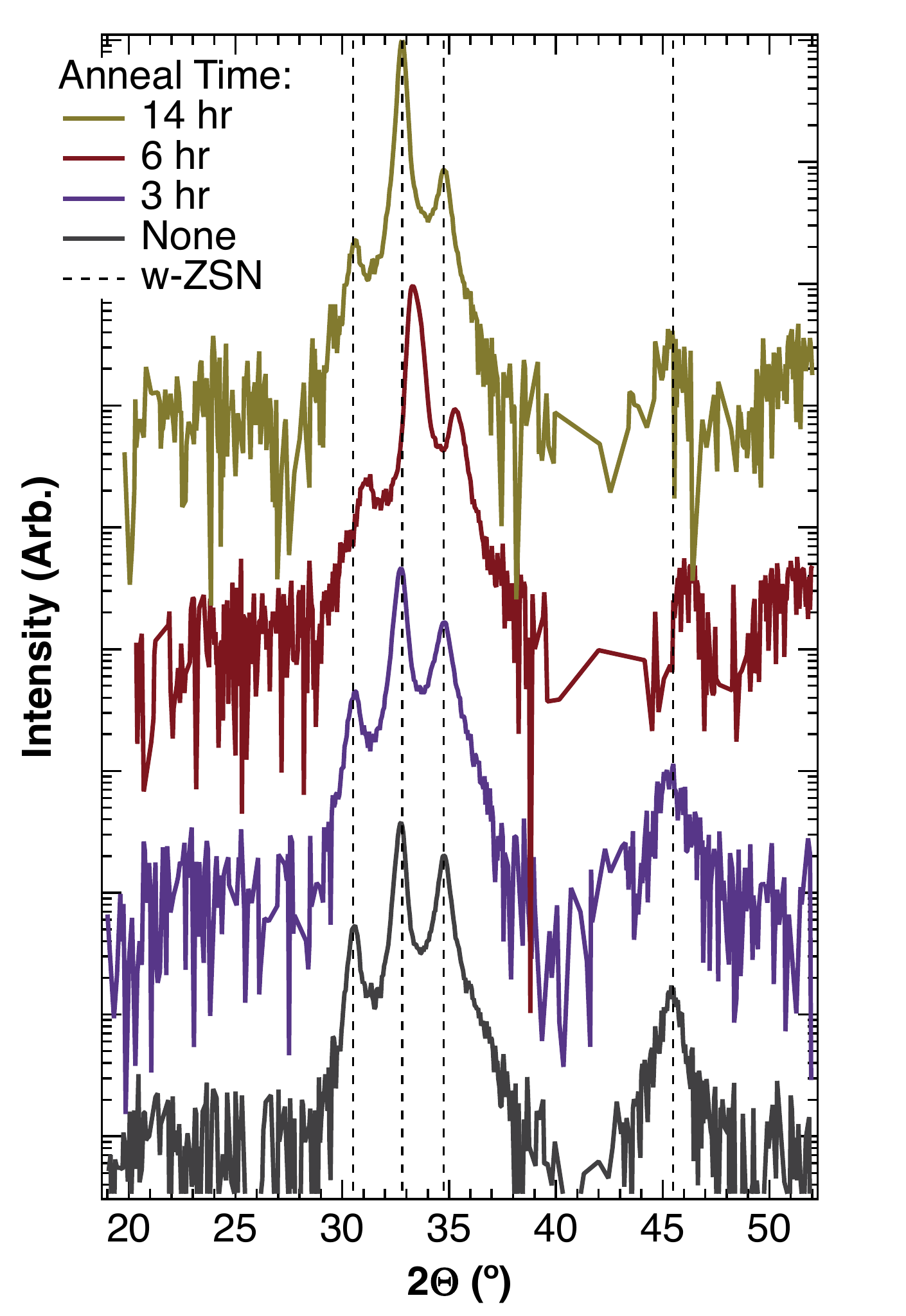}
	\caption{XRD patterns plotted on a log scale for ZnSnN\lo{2} films annealed for various times. The dashed vertical lines give the peak positions for wurtzite ZnSnN\lo{2}. Although preferential orientation along the c-axis grew more intense as annealing time increased (indicated by increasing intensity of the central (0001) peak), no deviation from the wurtzite structure was observed as a result of annealing treatment.}
	\label{fig:xrd}
\end{figure}

Only minimal changes in XRD pattern (Fig. \ref{fig:xrd}) were observed as films were annealed for increasing periods of time. All films described in this work exhibited XRD intensity consistent with wurtzite ZnSnN\lo{2} crystal structure with no evidence of super-lattice ordering (e.g. orthorhombic Pna2\lo{1} as observed in our previous work) \cite{fioretti2015}. Preferential orientation along the c-axis was encouraged by annealing treatment, as can be seen in Fig. \ref{fig:xrd}, particularly for the red and brown traces representing 6 hr and 14 hr annealing treatments, respectively. This result is somewhat promising, in that post-growth annealing can encourage uniaxial texturing in ZnSnN\lo{2}. However, the XRD data presented in Fig. \ref{fig:sem} along with the SEM images in Fig. \ref{fig:sem} collectively suggest that 300\degree C anneal temperature was too low to observe meaningful structural changes in ZnSnN\lo{2}. 

\begin{figure}[!h]
	\centering
		\includegraphics[width=6cm]{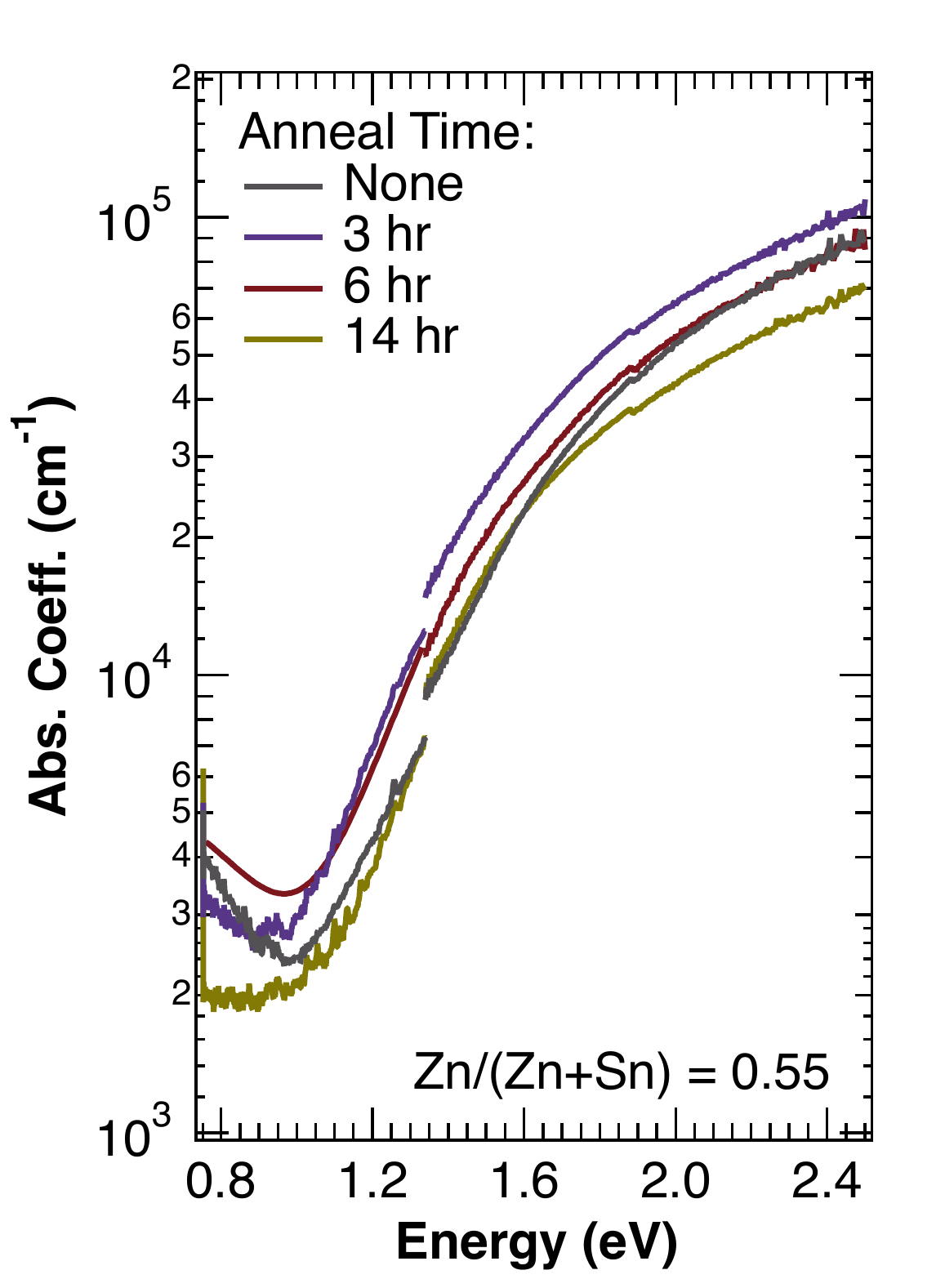}
	\caption{Absorption coefficient plotted on a log scale vs. photon energy for zinc-rich films annealed at 300\degree C for increasing time. No significant shift in absorption onset was observed as a function of annealing. Together with the findings from SEM and XRD, the absorption coefficient data plotted above likely indicates that 300\degree C annealing temperature did not provide enough thermal energy to induce the structural changes necessary to induce a shifting optical bandgap.}
	\label{fig:abs}
\end{figure}

The absorption coefficient data for zinc-rich Zn-Sn-N films annealed at 300\degree C for various times further supports the above conclusions pertaining to annealing temperature. No significant shift in the absorption onset was observed as annealing time increased for any of the films described in this work. Fig. \ref{fig:abs} shows absorption coefficient vs. photon energy for films with Zn/(Zn+Sn) = 0.55 plotted with respect to annealing time. The curve shape and absorption onsets displayed in Fig. \ref{fig:abs} are representative of stoichiometric films as well, with the exception that stoichiometric films exhibited more intense free carrier absorption below the absorption edge. 

The absence of a significant effect in absorption onset with annealing is consistent with the conclusion that annealing at 300\degree C did not provide enough thermal energy to enact the changes in crystal structure that would be expected to lead to a shifting absorption edge (i.e. inducing an ordered cation sublattice, which is tied to a higher-energy optical bandgap) \cite{feldberg2013}. In contrast to this explanation, another recent work on ZnSnN\lo{2} \cite{quayle2015} showed that sequential stacking of ordered orthorhombic unit cells of the types Pna2\lo{1} and Pmc2\lo{1} could produce enough structural disorder to satisfy entropic forces while still exhibiting wurtzite XRD and without the previously predicted decrease in bandgap \cite{chen2014}. If this were the case, no shift in absorption edge would be expected upon annealing. Whereas this latter explanation is intriguing, it does not line up with our previous finding that stoichiometric ZnSnN\lo{2} films grown at 280\degree C exhibited low-angle XRD peaks consistent with the Pna2\lo{1} orthorhombic structure previously reported \cite{lahourcade2013}. Given this inconsistency, and the agreement of the absorption coefficient data in Fig. \ref{fig:abs} with the findings from SEM and XRD, we conclude that it is more likely that annealing at 300\degree C did not provide sufficient thermal energy to induce significant shifts in the optical absorption onset.

\section{Conclusion}
\label{sec:conc}

In this work, thin films of compositionally-graded Zn-Sn-N were subjected to post-growth annealing treatment at 300\degree C for varying time to determine whether annealing would be a viable route toward improving the material properties for PV absorber applications. Quite non-intuitively, it was found that annealing treatment of tin-rich, stoichiometric, and zinc-rich films consistently lowered the mobility as a function of annealing time. This effect is likely indicative of non-octet-rule-preserving motif clustering, and highlights the need to grow ZnSnN\lo{2} films with a minimal amount of initial disorder before performing post-growth annealing treatments. Free electron concentration was found to decrease with annealing time up to 6 hr for stoichiometric and zinc-rich films, and increase again for samples annealed for 14 hr. Tin-rich films exhibited an increase in carrier density upon annealing treatment, which we propose should effectively rule out tin-rich Zn-Sn-N as a potential PV absorber material. Finally, post-growth annealing treatment did not produce noticeable changes in XRD pattern, absorption onset, or grain structure as a function of annealing time. This unexpected result likely indicates that annealing treatments above 300\degree C will be necessary to fully understand the effects of post-growth heat treatment on the properties of the Zn-Sn-N material system.


%



\section*{Acknowledgment}

This work was supported by the U.S. Department of Energy as a part of the Non-Proprietary Partnering Program under Contract No. De-AC36-08- GO28308 with the National Renewable Energy Laboratory. A.N.F. was supported by the Renewable Energy Materials Research Science and Engineering Center under Contract No. DMR-0820518 at the Colorado School of Mines. The authors would like to thank Dr. Andre Bikowski at the National Renewable Energy Laboratory for his technical assistance and ongoing maintenance of the combinatorial sputtering system used to produce the films in this work.




\newpage
\balance
\bibliographystyle{IEEEtran}
\bibliography{ZnSnN2}

\end{document}